\documentstyle[12pt]{article}
\def\be{\begin{equation}}
\def\ee{\end{equation}}
\newcommand{\zz}{%
\hspace{1mm}\begin{picture}(20,10)%
\put(0,0){\line(1,0){10}}%
\put(4.5,2){\mbox{\scriptsize{/}}}%
\end{picture}\hspace{-2.3mm}}
\newcommand{\cc}{-\circ -}

\begin{document}

\title{The Large Time
Behaviour in Quantum Field Theory and Quantum Chaos}
\author{
I.Ya. Aref'eva and I.V. Volovich
\\
{\it Steklov Mathematical
Institute, Russian Academy of Sciences}\\
{\it Gubkin St.8, GSP-1, 117966, Moscow, Russia}\\
{\it arefeva@mi.ras.ru, volovich@mi.ras.ru}
}
\date{}
\maketitle

\begin {abstract}
An exact general formula for the matrix elements of the evolution
operator in quantum theory is established. The formula
("$ABC$-formula") has the form $<U(t)>=\exp(At+B+C(t))$. The
constants $A$ and $B$ and the decreasing function $C(t)$ are
computed in perturbation theory.  The $ABC$-formula is valid for a
general class of Hamiltonians used in statistical physics and
quantum field theory. The formula includes the
higher order corrections to the well known  Weisskopf-Wigner
approximation and to the  stochastic (van Hove) limit which are
widely used in considerations of  problems of  radiation, decay,
quantum decoherence, derivation of master and
kinetic equations etc. The function $C(t)$ admits an
interpretation as an analogue of the autocorrelation function
describing quantum chaos for the quantum baker's map.
\end {abstract}
\newpage
\setcounter{equation}{0}


$~~~~~~~~~~~~~~~~~$ {\it Dedicated to Sergio Albeverio on the occasion of his 60th birthday}

\section{Introduction}

Scattering theory is one of the very old  and developed
topics in quantum theory.
The study of the large time behaviour of the evolution operator in
statistical physics and quantum field theory is the subject
of numerous investigations.
The basic object to study in quantum field theory is the scattering matrix,
see in particular \cite{Alb}--\cite{ARE}
and references therein.

However, there are many important problems in quantum field
theory  where we are interested  in the large but not infinite  time and
 where the standard $S$-matrix description is not
very convenient or
even not applicable. These include  processes with unstable
particles \cite{Sch,GW}, atom-photon interactions
\cite{CDG}, elementary particles in "semidressed states" with
non-equilibrium proper fields \cite{Fei}, electroweak baryogenesis and
phase transitions in the early Universe and in high-energy collisions
\cite{RS}, quantum optics \cite{WM},
quantum decoherence (see for example \cite{WM,Vol1})
 etc. In the consideration
of such processes we are interested in the time regime smaller
 than the "infinite" time when the $S$-matrix description
 becomes applicable.
The consideration of such processes
belongs to non-equilibrium quantum field theory, see
\cite{AAV} for more discussions.

Various approximate methods of consideration of time evolution for classical and
quantum systems have been developed by Bogoliubov \cite{Bog1},
Weisskopf and Wigner (see \cite{GW}), van Hove \cite{vHo1},
Prigogine \cite{PH} and many others, see \cite{AcKoVo,AcLuVo}.
The purpose of this paper is to obtain
an exact general result about the large time behaviour of certain matrix
elements
of the evolution operator

If $H$ is a generic self-adjoint operator in a Hilbert space  and we
are interested in the study of the corresponding evolution operator
$e^{-itH}$ then we can say nothing interesting about the behaviour
of its matrix elements
$$
(\psi,e^{-itH}\psi)=\int e^{-it\sigma}d\rho(\sigma)
$$
We have to restrict ourself to the Hamiltonians of some special forms if we want
to obtain useful results. In this paper we  consider
a  general class of  Hamiltonians used in solid
state physics and quantum field theory
\begin{equation}
H=H_0+\lambda V, \label{1}
\end{equation}
where $H_0$ is a free
Hamiltonian,  $V$ describes an interaction and $\lambda$
is the coupling constant. We shall study the evolution operator
\begin{equation}
U(t)=e^{itH_0}e^{-itH}.\label{2}
\end{equation}
The main result of this paper is the following exact  formula
(we call it the $ABC$-formula) which is valid for arbitrary
time $t$
\begin{equation}
\langle U(t)\rangle=e^{At+B+C(t)}.
\label{I4}
\end{equation}
Here $A$ and $B$ are constants for which a representation
  in perturbation theory will be given and
$C(t)$ is a function which under rather general assumptions
 can be represented for large time $t$ as
 \begin{equation}
C(t)={f(t)\over t^{\alpha}}.
\label{5}
\end{equation}
Here
$f(t)$ is a bounded  function
and  the exponent $\alpha$ depends on the model and
on the dimension of space ($\alpha=3/2$ for the physical
3-dimensional space). The function $C(t)$ admits a representation which has
the form similar to the autocorrelation function describibg quantum chaos,
see below.

The well known Weisskopf-Wigner
approximation \cite{GW}
is given by
\begin{equation}
\langle U(t)\rangle\simeq e^{at},
\label{3}
\end{equation}
where $a$ is a constant. It corresponds to the particular
case of the $ABC$ - formula (\ref{I4}).

Expectation value in (\ref{I4}) is taken over the vacuum vector. For the case
of one-particle states we obtain
\begin{equation}
\langle p| U(t)|p'\rangle=e^{iA(p)t+B(p)+C(t,p)}\delta (p-p').
\label{I4a}
\end{equation}

The formulae (\ref{I4}) and  (\ref{I4a}) have a very general character.
We prove (\ref{I4}) in
Section 3 for a wide class of Hamiltonians.
The class of considered Hamiltonians includes the Bose and Fermi gases,
phonon self-interaction and electron-phonon interaction, quantum
electrodynamics in external fields etc.

We derive the main formula (\ref{I4}) by using the theory of perturbation
of  spectra and renormalized wave
operators \cite{Hepp,ARE}. This method can be used only under some restrictions
to the form of the Hamiltonian when
one has not decay. Another method based on the  direct examination of
perturbation
theory which  can be used
also  in the case of decay will be considered in another publication, see \cite{AV5}.

\section{Notations and auxiliary results }
\subsection{Hamiltonians}

 We consider
Hamiltonians of the  form (\ref{1}) where $H_0$ is a free
Hamiltonian
\begin{equation}
\label{I1}
H_0=\sum _i \int \omega _i(k)a^*_i(k)a_i(k)d^dk
\end{equation}
and $V$ is the sum of Wick monomials.
Creation and annihilation operators $a^*_i(k)$, $a_i(k)$
describe particles or quasiparticles and they
satisfy the commutation or anticommutation relations
\begin{equation}
\label{I2} [a_i(k), a^*_j(k')]_{\pm}=\delta _{ij}\delta (k-k')
\end{equation}
Here $k,k^{'}\in R^d$ and $i,j=1,..,N$ label the finite
number of different types of (quasi)particles. Creation and annihilation
operators have the standard realization in the Fock space
with the vacuum vector which will be denoted $|0\rangle$ or $\Phi_0$ .
Examples of one-particle energy $\omega _i(k)$
include the relativistic ($\omega (k)=(k^2+m^2)^{1/2}$)
and non-relativistic ($\omega (k)= k^2/2-\omega_0$) laws, the Bogoliubov
spectrum ($\omega (k)= (bk^4+k^2v(k))^{1/2}$), the Fermi quasiparticle
spectrum ($\omega(k)=|k^2/2m-\mu|$) etc.

 We consider two
different types of Wick polynomials.  The first type describes an
interaction in the case when there is not the  translation invariance

\begin{equation}
V=\sum_{I,J,i,j }
\int v(p_1,i_1\dots p_{I},i_I|q_1,j_1\dots q_J,j_J)
\prod^I_{l=1}a^*_{i_l}(p_l) dp_l\prod^J_{r=1}a_{j_r}(q_r)dq_r \label{4}
\end{equation}
were $v(p_1,i_1,\dots p_{I},i_I|q_1,\dots q_J,j_J)$ are some test  functions.

The second type is described by the translation invariant
Hamiltonian

\begin{equation}
V=\sum_{I,J}V_{I,J}=\sum_{I,J,i,j}\int \hat {v}
(p_1,i_1,\dots p_{I},i_I|q_1,j_1\dots
q_J,j_J)\label{H8}
\end{equation}

$$
\delta
\left(\sum^I_lp_l-\sum^J_rq_r\right)\prod^I_{l=1}a^*_{i_l}(p_l)
dp_l\prod^J_{r=1}a_{j_r}(q_r)dq_r
$$

Clearly the delta function causes the trouble and there are  singular
terms in (\ref{H8}). Namely, $V_{I,0}\phi $ does not belong to the Fock
space unless the vector $\phi =0$.
This singularity is called the volume singularity. To give a meaning
to the Hamiltonian  with interaction (\ref{H8})
one has to introduce a volume
cut-off, then perform the vacuum renormalization and
vacuum dressing and only after that
remove the cut-off. This procedure defines the Hamiltonian
in a  new  space (see \cite{Hepp,Are} for details). To avoid this
difficulty in this paper we  will assume that for translation
invariant interaction there are no pure creation and annihilation terms.

\subsection{Friedrichs diagrams }

We will study  the evolution operator
\begin{equation}
\label{U1a}
U(t)=e^{itH_0}e^{-it(H_0+\lambda V)}.
\end{equation}
 In  perturbation theory the evolution operator (\ref{U1a})
has the representation
\begin{equation}
U(t)=1-i\lambda\int^t_0V(t_1)dt_1+
(-i\lambda)^2\int^t_0dt_1\int^{t_1}_0dt_2V(t_1)
V(t_2)+\dots
\label{U2}
\end{equation}
where
$$
V(t)=e^{itH_0}Ve^{-itH_0}
$$

We will use for $V_{I,J}$ the Friedrichs diagram representation \cite{Frie}.
The corresponding diagram has one vertex and $I$ lines
going from the vertex
to the left and $J$ lines going to the right.
The first $I$ lines represent creation operators and the last $J$
lines represent annihilation operators. In what follows we will use
the Wick theorem and also the following notions.
The line of the graph is called the internal if it connects two vertices
of the graph.
A graph is the connected graph if
all its vertices are connected by a set of
internal lines otherwise it is called the  disconnected one.
A connected graph is called the one-particle reducible (1PR)
if after the removing a line it becomes  disconnected.
A connected graph is called the one-particle irreducible (1PI)
if after the removing any line it is still connected.

The following "linked cluster theorem" \cite{Hepp} will be used:

\begin{equation}
\label{1a}
U(t)=:e^{U_c(t)}\ :
\end{equation}
where
$$
U_c(t)=
\sum_{n=1}^{\infty}
(-i\lambda)^n\int^t_0dt_1...\int_0^{t_{n-1}}dt_n
\left(V(t_1)...V(t_n)\right)_c
$$
Here $::$ means the Wick normal ordering and the index $c$  in $U_c$
indicates that one takes only the
connected diagrams.

Below  for simplicity of notations  we consider interactions with
the only one type of particles, but the main results are valid for
arbitrary number of types of particles.

\section { Non-translation invariant Hamiltonians}
\subsection{Second order}
To get an insight to the problem it is very instructive to start with the
consideration of
matrix elements of the evolution operator in the second order of
perturbation theory.
For the vacuum matrix element of
the evolution operator we  obtain from (\ref{1a}) the
representation:
\begin{equation}
\langle 0|U(t)|0\rangle=e^{{\cal E}(t)}
\label{L1}
\end{equation}
where
\begin{equation}
{\cal E}(t)=\langle 0|(-i\lambda\int^t_0dt_1V(t)+(-
i\lambda)^2\int^t_0dt_1\int^{t_1}_0dt_2V(t_1)V(t_2)+\dots)_c|0\rangle .
\label{L2}
\end{equation}
Here the symbol $(...)_c$ means that we keep only connected diagrams.

Representation (\ref{L1}), (\ref{L2})
permits us to calculate the leading terms of the asymptotic behaviour
of the matrix elements of the evolution operator
for large time $t$
as well the corrections to the leading terms.
In fact we will show that  ${\cal E}(t)$
has the following form
\begin{equation}
{\cal E}(t)=At+B+C(t)
\label{E2}
\end{equation}
where one has the perturbative expansions
\begin{equation}
\label{EX1}
A=\lambda^2A_2+\lambda^3A_3+..., ~~~B=\lambda^2B_2+\lambda^3B_3+...,~~~
C(t)=\lambda^2C_2(t)+\lambda^3C_3(t)+...
\end{equation}
and $C_n(t)$ vanishes for large $t$.

Let us find explicitly these terms in the
second order of perturbation theory
for the Hamiltonian

\begin{equation}
H=H_0+\lambda V, \label{H1}
\end{equation}
where
\begin{equation}
\label{H2}
H_0=\int \omega (p)a^*(p)a(p)dp
\end{equation}
and the interaction has the form
\begin{equation}
V=\int (v(p_1,..., p_n)a^*(p_1)...a^*(p_n)+h.c.).
dp_1...dp_n
\label{3.30}
\end{equation}
Here $\omega (p)$ is a positive smooth function, for example
$\omega(p)=\sqrt{p^2+m^2}, m>0$
and $v(p_1,...,p_n)$ is a   test  function.
For this interaction the first term in
(\ref{L2}) is identically zero.
The second term in (\ref{L2}) equals to

\begin{equation}
           \label{3.2}
{\cal E}^{(2)}(t)
= (-i\lambda)^2\int dp_1...dp_n |v(p_1,...,p_n)|^2
\int^t_0dt_1\int^{t_1}_0dt_2
e^{it_1E_1+it_2E_2}
\end{equation}
where
\begin{equation}
\label{3.3}
E_2=-E_1=E(p_1,...,p_n)=\sum_{i=1}^{n}\omega(p_i)
\end{equation}

By using the equality
\begin{equation}
\label{Exp1}
\int^t_0dt_1\int^{t_1}_0dt_2
e^{-it_1E+it_2E}=
-\frac{i}{E}t+\frac{1}{E^2}-\frac{1}{E^2}e^{-itE}
\end{equation}
we get
\begin{equation}
\label{EXP3}
{\cal E}^{(2)}(t)=
(-i\lambda)^2\int dp_1...dp_n |v(p_1,...,p_n)|^2
(-\frac{i}{E}t+\frac{1}{E^2}-\frac{1}{E^2}e^{-itE})
\end{equation}

Therefore we obtain the expression of the form (\ref{E2})
\begin{equation}
\label{EXP4}
{\cal E}(t)=\lambda^2A_2t+\lambda^2B_2+\lambda^2C_2(t)+...
\end{equation}
where
\begin{equation}
\label{A1_2}
A_2=i\int
\frac{|v(p_1,...,p_n)|^2}{E(p_1,...,p_n)}dp_1...dp_n,
\end{equation}

\begin{equation}
\label{T2_2}
B_2=-\int
\frac{|v(p_1,...,p_n)|^2}{E(p_1,...,p_n)^2}dp_1...dp_n,
\end{equation}

\begin{equation}
\label{T3_2}
C_2(t)=\int
\frac{|v(p_1,...,p_n)|^2}{E(p_1,...,p_n)^2}
e^{-itE(p_1,...,p_n)}dp_1...dp_n
\end{equation}

We have obtained the following

{\bf Theorem 1.}
 The vacuum expectation value of the evolution operator for
the Hamiltonian (\ref{H1}) in the second order of perturbation
theory has the form
\begin{equation}
\label{as2}
<U(t)>=e^{\lambda^2A_2t+\lambda^2B_2+\lambda^2C_2(t)}
\end{equation}
where $A_2,B_2$ and $C_2(t)$ are given by (\ref{A1_2}),(\ref{T2_2})
and (\ref{T3_2}).

{\bf Remark}.  By using
the stationary phase method one can prove that the function
$C_2(t)$  vanishes as $t\to\infty$ (see below).

\subsection{Decay}

We have proved theorem 1 under the assumption $\omega(p)>0$.
However the obtained formula (\ref{as2}) is valid in the
more general case when one has the decay.
In this case formula (\ref{as2}) still is true but
in the expressions (\ref{A1_2})-(\ref{T3_2}) one
has to substitute $E\to E-i0$. Let us consider the important case
when
\begin{equation}
\omega(p)={p^2\over 2}-\omega_0,~~\omega_0>0
\label{omega}
\end{equation}
Instead of (\ref{Exp1}) we will use now the identity

\begin{equation}
\label{Exp12}
\int^t_0dt_1\int^{t_1}_0dt_2
e^{i(t_2-t_1)E}=
t\int_0^t(1-{\sigma\over t})e^{-i\sigma E}d\sigma .
\end{equation}
We have
\begin{equation}
\label{3.2S}
{\cal E}^{(2)}(t)=
(-i\lambda)^2\int dp_1...dp_n |v(p_1,...,p_n)|^2
\int^t_0dt_1\int^{t_1}_0dt_2
e^{i(t_2-t_1)E(p_1,...,p_n)}
\end{equation}
$$
=-\lambda^2 t\int_0^td\sigma(1-{\sigma\over t})
\int dp_1...dp_n
|v(p_1,...,p_n)|^2e^{-i\sigma E(p_1,...,p_n)}
$$
$$
=\lambda^2 tA_2(t)+\lambda^2B_2(t)
$$
where
\begin{equation}
\label{3.2T}
A_2(t)=
-\int_0^td\sigma F(\sigma),~~B_2(t)=
\int_0^td\sigma\sigma F(\sigma)
\end{equation}
and
\begin{equation}
\label{3.3TR}
F(\sigma)=
\int dp_1...dp_n
|v(p_1,...,p_n)|^2e^{-i\sigma E(p_1,...,p_n)}.
\end{equation}
By using the stationary phase method we obtain the following
asymptotic behaviour of the function $F(\sigma)$ as
$\sigma\to\infty$:
$$
F(\sigma)=({2\pi i\over\sigma})^{dn/2}e^{in\sigma\omega_0}
|v(0)|^2[1+o({1\over\sigma})].
$$
Therefore for $dn\geq 3$ there exist the limits
\begin{equation}
\label{3.3TP}
\lim_{t\to\infty}A_2(t)=A_2=-\int_0^{\infty}d\sigma F(\sigma)
\end{equation}
$$
\lim_{t\to\infty}B_2(t)=B_2=\int_0^{\infty}d\sigma
\sigma F(\sigma)
$$
because there exists the limit
$$
\lim_{t\to\infty}\int_1^{t}e^{i\sigma\omega_0}{d\sigma\over
\sigma^{1/2}}.
$$
Moreover one has
$$
A_2(t)=-\int_0^{\infty}d\sigma F(\sigma)+o({1\over t^2}),~~
B_2(t)=\int_0^{\infty}d\sigma
\sigma F(\sigma)+o({1\over t}).$$
If $dn\geq 5$ one gets also
\begin{equation}
\label{3.3TI}
A_2=i\int dp_1...dp_n
{|v(p_1,...,p_n)|^2\over E(p_1,...,p_n)-i0}
\end{equation}

\begin{equation}
\label{3.3TG}
B_2=-\int dp_1...dp_n
{|v(p_1,...,p_n)|^2\over (E(p_1,...,p_n)-i0)^2}.
\end{equation}
Indeed one has
$$
A_2=-\int_0^{\infty}d\sigma F(\sigma)=
-\lim_{\epsilon\to 0}\int_0^{\infty}d\sigma
F(\sigma)e^{-\sigma \epsilon}.
$$
This is true due to the Lebesgue theorem since
$|F(\sigma)e^{-\sigma \epsilon}|\leq |F(\sigma)|$
and $F(\sigma)\in L_1(R_+)$ ($L_1$ is the  space of absolute integrable
functions) if $nd\geq 3$. Substituting in the above formula the
representation (\ref{3.3TR}) and changing the order of integrations
(we can do this
due to the Fubini theorem since for positive $\epsilon$
the function $|v(p_1,...,p_n)|^2e^{-i\sigma (E(p_1,...,p_n)-i\epsilon)}$
 belongs to the space $L_1(R_+\times R^{nd})$ of
absolute integrable functions), we can
perform the integration over $\sigma$ explicitly
$$A_2=\lim_{\epsilon\to 0}(i)
\int dp_1...dp_n
{|v(p_1,...,p_n)|^2\over E(p_1,...,p_n)-i\epsilon}
=i\int dp_1...dp_n
{|v(p_1,...,p_n)|^2\over E(p_1,...,p_n)-i0}.
$$
 The same calculation is true for $B_2$ with the more strong
 assumption : $dn\geq 5$,
$$
B_2=\int_0^{\infty}d\sigma\sigma F(\sigma)
=\lim_{\epsilon\to 0}\int_0^{\infty}d\sigma\sigma
\int dp_1...dp_n
|v(p_1,...,p_n)|^2e^{-i\sigma (E(p_1,...,p_n)-i\epsilon)}
$$
$$
=-\lim_{\epsilon\to 0}
\int dp_1...dp_n
{|v(p_1,...,p_n)|^2\over (E(p_1,...,p_n)-i\epsilon)^2}
=-\int dp_1...dp_n
{|v(p_1,...,p_n)|^2\over (E(p_1,...,p_n)-i0)^2}.
$$
We have proved the following theorem.

{\bf Theorem 2.}
 The asymptotic behaviour as $t\to\infty$ of the
vacuum expectation value of the evolution operator for
the Hamiltonian (\ref{H1})  with the dispersion low
(\ref{omega})
in the second order of perturbation
theory is
\begin{equation}
\label{as32}
<U(t)>=e^{\lambda^2A_2t+\lambda^2B_2+\lambda^2 o(1/t)}
\end{equation}
where $A_2$ and $B_2$ are given by (\ref{3.3TP})
(or (\ref{3.3TI}) and (\ref{3.3TG})).
After the rescaling $t\to t/\lambda^2$ one gets the $\lambda^2$
corrections to the stochastic limit
\begin{equation}
\label{as33}
<U(t/\lambda^2)>=e^{A_2t+\lambda^2B_2+\lambda^2 o(\lambda^2/t)}.
\end{equation}

\subsection{Example}

We discuss here the evolution operator for the
simple explicitly solvable model described by  the Hamiltonian
\begin{equation}
H=\int\omega(k)a^+(k)a(k)d^d k+\lambda\int(a(k)\overline v(k)+
a^*(k)v(k))d^dk.
\label{(1)}
\end{equation}
We will see that the vacuum expectation value of the
evolution operator has the form obtained in
theorems  and 2.
Under assumptions
\begin{equation}
\label{(2)}
 A=\lambda^2A_2=i\lambda^2\int
{|v(k)|^2\over\omega(k)}\,d^dk<\infty,~\quad
B=\lambda^2B_2=-\lambda^2\int{|v(k)|^2\over\omega^2(k)}\,d^dk<\infty\ ,
\end{equation}
one has the following

{\bf Proposition 1.}
The vacuum expectation value of the
evolution operator  $U(t)=e^{itH_0}e^{-itH}$
for the model (\ref{(1)}) is
\begin{equation}
\langle U(t)\rangle=\exp\left[At+
B+C(t)\right]
\label{(3)}
\end{equation}
where $A$ and $B$ are given by (\ref{(2)}) and $C(t)$ is
$$
C(t)=\lambda^2\int dk{|v(k)|^2\over\omega^2(k)}\,
e^{-i\omega(k)t}
$$

{\it Proof.} It follows from the known explicit solution of the
model.

$~$ From Proposition 1 we obtain

{\bf Proposition 2.}
 The asymptotic behaviour of the
expectation value (\ref{(3)}) for $t\to\infty$ has the form
\begin{equation}
\label{(4)}
\langle U(t)\rangle=\exp\left[At+
B+\lambda^2\left({1\over t}\right)^{d\over2}\,(2i\pi)^{d
\over2}\frac{|v(k_0)|^2}{\omega^2(k_0)}\,e^{-i\omega(k_0)t}+...\right].
\end{equation}
where $k_0$ is a critical point, $\nabla\omega(k_0)=0$ and we assume
there is only one nondegenerate critical point.

{\it Proof.} It follows immediately from (\ref{(3)})
by using the
stationary phase  method.

{\bf Remark}. If $\omega(k)=k^2-\omega_0,~\omega_0 >0 $
then one has the decay.
We can not use in this case the diagonalization of the Hamiltonian
(\ref{(1)}) but the formula (\ref{L1}) still is true. We have
$$
<U(t)>=\exp[- \lambda^2\int_0^tdt_1\int_0^{t_1}dt_2\int d^dk
|v(k)|^2e^{i(t_2-t_1)\omega(k)}]$$
$$
=\exp[\lambda^2tA_2(t)+\lambda^2B_2(t)]
$$
where
$$
A_2(t)=-\int_0^td\sigma F(\sigma),
~~B_2(t)=\int_0^td\sigma\sigma F(\sigma)
$$
and
$$
F(\sigma)=\int d^dk
|v(k)|^2e^{-i\sigma\omega(k)}
$$
Again we obtain the $ABC$-formula.

\subsection{Wave operators and the main formula}

In this section we show how the spectral theory and renormalized
wave operators  can be used for the derivation
of the main formula.
In particular explicit expressions for the
parameters $A,B$ and $C(t)$
will be obtained.

We consider the Hamiltonian (\ref{4}) for one type of particles
with $\omega(p)=~~~~~$ $\sqrt{p^2+m^2}$, $m>0$
in the space $R^d, d>2$. We will work with the formal perturbation
series for the evolution operator. In fact if the interaction
(\ref{4}) includes only fermionic operators or it is linear
in bosonic operators then one can prove that the series are absolutely
convergent. The following operator plays the crucial role in the
scattering theory
\begin{equation}
\label{Tfor}
T=:\exp\sum_{n=1}^{\infty}(-\lambda)^n
\Gamma(V...\Gamma(V)...)_L:
\end{equation}
Here $\Gamma$ means  the  Friedrichs
$\Gamma $ - operation
\begin{equation}
\label{H7}
\Gamma (V)=\lim _{\epsilon \to +0}(-i)\int _0^{\infty}
e^{-\epsilon t}e^{itH_0}Ve^{-itH_0}dt
\end{equation}
and $()_L$ means that only connected non-vacuum
diagrams are included.
The operator $T$ is equal in fact to the non-vacuum part
of the conjugate wave operator:
$$
T=\lim_{\epsilon\to 0}\lim_{t\to\infty}
U^*_{\epsilon}(t)/<U^*_{\epsilon}(t)>
$$
One has the following relations
\cite{Hepp}:
\begin{equation}
\label{Diag}
HT=T(H_0+E_0),
\end{equation}
$$
E_0=\sum_{n=1}^{\infty}(-\lambda)^{n+1}
<V\Gamma(V...\Gamma(V)...)_c>,
$$
$$T^*T=TT^*=Z^{-1},
$$
$$
Z^{-1}=||T\Phi_0||^2
$$
We will use these relations to derive the
main formula for matrix elements
of the evolution operator and in
particular to compute corrections
to the stochastic limit.

$~$From (\ref{Diag}) it follows
$$
H=T(H_0+E_0)T^*e^B
$$
where
\begin{equation}
\label{Bfor}
B=\ln Z
\end{equation}
Therefore one has
\begin{equation}
\label{U(t)}
U(t)=e^{itH_0}e^{-itH}=e^{itH_0}Te^{-it(H_0+E_0)}T^*e^B
=e^{At+B}e^{itH_0}Te^{-itH_0}T^*
\end{equation}
where
\begin{equation}
\label{Afor}
A=-iE_0
\end{equation}
By taking the expectation value of the equality (\ref{U(t)})
we obtain
\begin{equation}
\label{psi1}
<\psi,U(t)\psi>=e^{At+B+C(t)}
\end{equation}
where
$$
e^{C(t)}=<\psi,e^{itH_0}Te^{-itH_0}T^*\psi>
$$
If $\psi$ is the vacuum vector then one can prove that $C(t)\to 0$ as
$t\to\infty$ and we obtain the main formula (\ref{I4}). If $\psi$
is a non-vacuum vector then the asymptotic behaviour of $C(t)$
is more complicated. We have proved the following theorem.

{\bf Theorem 3.}
 If the Hamiltonian satisfies
the indicated above assumptions
then there exists the following
representation for the vacuum expectation
value of the evolution operator
$$
<\Phi_0,U(t)\Phi_0>=e^{At+B+C(t)}
$$
where constants $A$ and $B$ are given by (\ref{Afor}) and (\ref{Bfor})
and $C(t)$ is defined by
\begin{equation}
e^{C(t)}=<\Phi_0,T(t)T^*\Phi_0>,~~
T(t)=e^{itH_0}Te^{-itH_0}
\label{LC}
\end{equation}
Here the weak limit of $T(t)$ as $t \to \infty$
is equal to 1 and $\lim_{t\to\infty}C(t)=0$.

This theorem also shows
the physical meaning of constants $A,B$ and the function $C(t)$.

\section {One particle  matrix elements of the evolution operator for
translation invariant Hamiltonian}
In this section we study the asymptotic behaviour of one
particle matrix elements of evolution operator

\begin{equation}
\label{U1}
<p|U(t,\lambda)|p'>= \delta(p-p')U_{1,1}(t,p,\lambda),
\end{equation}
for translation invariant Hamiltonian (\ref{H8}) without
vacuum polarization, i.e. when $V_{I0}=V_{0J}=0$, and we assume also
$V_{1,1}=0$.
We will prove that the following

esentation is true
\begin{equation}
U_{1,1}(t,p,\lambda)=
\exp \{ itA(p,\lambda)+B(p,\lambda)\}\left(1+C(t,p,\lambda)\right)
\label{N7a}
\end{equation}
where $A(p,\lambda)$, $B(p,\lambda)$ and $C(t,p,\lambda)$
are formal series in $\lambda$
\begin{equation}
A(p,\lambda) =\sum _{n=1}^{\infty}\lambda^{n}A_n(p),~~
B(p,\lambda) =\sum _{n=1}^{\infty}\lambda^{n}B_n(p),~~
C(t,p,\lambda) =\sum _{n=1}^{\infty}\lambda^{n}C_n(t,p)
\end{equation}
and functions  $C_n(t,p)$ vanish as $t \to \infty$.

We show how the spectral theory and renormalized wave
operators  can be used for
the derivation
of the main formula.
In particular, explicit recursive  relations
for the parameters $A_n,B_n$ and $C_n(t)$
will be obtained. Note that for this derivation
we have to assume that the is no decay.

The intertwining operator $T$ is defined as a
solution of the following equation
\begin{equation}
\label{Inter}
HT=T(H_0+M),
\end{equation}
Here $M$ has the form
\begin{equation}
\label{mex}
M=\int m(p)a^*(p)a(p)dp
\end{equation}
The operator $T$ plays the crucial role in the scattering
theory. Its singular part defines the renormalized wave operators.
The renormalized wave operators
also give a solution of intertwining condition.
Taking $T$ in the following form
\begin{equation}
\label{TW}
T=:\exp W:, ~~~W=\Gamma(Q)
\end{equation}
one gets \cite{ARE}   equations to define $Q$ and $M$
\begin{equation}
\label{T}
Q+V\zz T-(V\zz T)_{1,1}-W\cc M=0
\end{equation}
\begin{equation}
\label{M}
M=(V\zz T)_{1,1}
\end{equation}
Here the symbol $\zz$ means that for connected $A$ in the
operators $A\zz B$  all connected parts
of B are paired with $A$. If $B$ is connected then $A\zz B=
A\cc B=(AB)_c$.
For a special form of the interaction, when $V_{0I}=V_{I0}=0$
we can write  $M=(V\zz T)_{1,1}=(V\Gamma(Q))_{1,1}$.

Expanding $M$ and $Q$ in the power series in $\lambda$
\begin{equation}
\label{se}
M=\sum _{n=1}^{\infty} \lambda ^{2n} M_{2n},~~
Q=\sum _{n=1}^{\infty} \lambda ^{2n+1} Q_{2n+1},
\end{equation}
we get recursive relations to define $M_{2n}$ and $Q_{2n+1}$.
Let us compute explicitly the first terms solving these equations.
We obtain
\begin{equation}
\label{Q1}Q_1=-V
\end{equation}
\begin{equation}
\label{M2}
M_2=-(V\Gamma V)_{1,1}
\end{equation}
\begin{equation}
\label{Q2}
Q_2=(V\Gamma V)_{c}-(V\Gamma V)_{1,1}
\end{equation}
\begin{equation}
\label{Q3}
Q_3=
-(V\Gamma _r(V \Gamma (V)))_c-\frac{1}{2}V
\zz :\Gamma (V)^2:+
\Gamma V\cc M_{2}
\end{equation}
\begin{equation}
\label{M4}
M_4=
-(V\Gamma(V\Gamma _r(V \Gamma (V))))_{1,1}+
(V\Gamma ^2(V)M_{2})_{1,1}
\end{equation}
or
\begin{equation}
\label{M42}
M_4=
-(V\Gamma(V\Gamma _r(V \Gamma (V))))_{1,1}
-(V\Gamma ^2(V)(V \Gamma (V))_{1,1})_{1,1}
\end{equation}

Here we use the notation
\begin{equation}
\label{rg}
\Gamma _r(Q)=\Gamma (Q-Q_{1,1})
\end{equation}
One can  construct in perturbation theory
the operator $Z$ such that
\begin{equation}
\label{T*}
TT^*Z=1
\end{equation}
In particular, at the second order of perturbation
theory in $\lambda$ one has
\begin{equation}
<p|Z|p'>=(1 + \lambda ^2B_2(p))\delta (p-p')
\end{equation}
By using (\ref{Inter}) and (\ref{T*}) one gets
\begin{equation}
\label{U(t)11}
U(t)=e^{itH_0}e^{-itH}=e^{-itM}e^{it(H_0+M)}
Te^{-it(H_0+M)}T^*Z
\end{equation}
By taking the one particle expectation value of the
equality (\ref{U(t)11}) we obtain
\begin{equation}
\label{ABc}
<p|U(t)|p'>=e^{-iM(p,\lambda )t}Z(p,\lambda )(1+C(p,t, \lambda ))
\end{equation}
where
\begin{equation}
(1+C(t,p,\lambda )) \delta (p-p')=<p|e^{it(H_0+M)}Te^{-it(H_0+M)}T^*|p'>
\label{Ct}\end{equation}
and $M$ and $T$ should be computed from the recursive relations.

We have proved the following

{\bf Theorem 4.}
 For translation invariant Hamiltonians without vacuum
polarization the one particle matrix elements of evolution operator
are given by the formula
(\ref{ABc}) where functions $M(p,\lambda )$ and $Z(p,\lambda )$
are solutions of equations (\ref{TW})-(\ref{M}) and (\ref{T*}).
Function $C(t,p,\lambda )$ is defined by (\ref{Ct})
and it vanishes as $t \to \infty$.

\section{Quantum Chaos}

The representations (\ref{LC}) and (\ref{Ct}) for the function $C(t)$
are similar in fact to the representation for the autocorrelation function
describing quantum chaos for the quantum baker's map.
The quantum baker's map
is a simple model invented to study quantum chaos, see for example
\cite {BV,TVS,KH,SS,IOV} and references therein.  The evolution operator
for the quantum baker`s map has the form $U(t)=B^t$ where $B$ is a unitary
matrix and
$t$ is an integer number. The autocorrelation function
$F(t)=<\psi|U(t)|\psi>$ where
$\psi$
is a coherent vector, admits a representation of the form \cite{KH,TVS}
$$
F(t)=a(t)\sum e^{W(\beta,p,q,t)}
$$
where $a(t)$ falls off exponentially for large $t$ and $W$ is
quadratic polynomial in $q,p$ and $\beta$. Here $p$ and $q$ are
parameters in the coherent vectors, $\beta $ is a
string in the symbolic dynamics representation and the sum runs
over  all the strings. This form is similar to the representations
(\ref{LC}) and (\ref{Ct}) for the function $C(t)$, compare  also with the
explicit
formula (\ref{T3_2}) for $C_2(t)$. It would be very interesting to study
further this analogy between the functions $F(t)$ and $C(t)$. In particular
one has to investigate
the sensitivity of  the function $C(t)$ to the initial data, i.e. in the
one particle case
the dependence of the function $C(t,p,\lambda)$ on $p$. Also one has to
investigate
the analogous function for coherent states. To this end one can use the
perturbation theory
as well the semiclassical expansion. Actually it is possible that a quantum system 
exhibits chaotic properties, although its classical counterpart is non-chaotic.
Such phenomena are called wave chaos \cite{AlbSeb}. It would be very interesting
to study a relation of the properties of the function $C(t)$
with the wave chaos  and also with  the spectral
and holographic properties of quantum chaos, see \cite{AMRV}.

\bigskip

\section{Conclusion}

We have obtained in this paper the explicit representations (\ref{I4})
for the vacuum and one-particle matrix elements of the evolution operator.
By using these representations we have computed the corrections to the
known results for the large time exponential behaviour
of these matrix elements. This opens the way for
further investigations of the large time behaviour in
quantum theory. In particular the problems of quantum decoherence
and decay and also the relation with quantum chaos
require the further study.

{\bf Acknowledgements.} This work was supported in part by RFFI-99-01-00166 (I.Ya.A.)
and  by RFFI-99-01-00105 (I.V.V.) and by grant for the leading
scientific schools 96-15-96208 (I.Ya.A.) and 96-15-96131 (I.V.V.).


\begin{thebibliography}{99}
\bibitem{Alb}
S.\ Albeverio , D.\ Bolle , F.\ Gesztesy , R.\ Hoegh-Krohn  and L.\ Streit,
{\it Low-energy parameters in nonrelativistic
scattering theory},  Ann. Phys., New York,
 \  148 (1983), 308--337.

\bibitem{BS}
N.~N. \ Bogoliubov and  D.~V. \ Shirkov,
{\it Introduction to the theory of quantum fields}, Nauka, Moscow, 1973.

\bibitem{Frie}
K.~O. \ Friedrichs, { \it Perturbation of  Spectra
in Hilbert Space}, AMS, Providence, 1965.

\bibitem{LD}
L.~D. \ Faddeev, {\it Separation of selfinteraction effects
and scattring in the perturbation theory},
 Dokladi AN USSR, \  152 (1963), 573--578.

\bibitem{Hepp}
K. \ Hepp, {\it  Theorie de la renormalisation},
Springer, New York, 1969.

\bibitem{ARE}
I.~Ya. \ Aref'eva, {\it Renormalized waves operators, I},
\ Teor. \ Mat. \ Fis. \  14 (1973),
3--17.

\bibitem{Sch}
J. \ Schwinger, {\it Field theory of unstable particles},
Ann. Phys., New York, \  9 (1960), 169-193

\bibitem{GW}
M.~L.\ Goldberger and K.~M. \ Watson, \
 {\it  Collision theory},
John Wiley \& Sons, Inc, New York-London-Sydney, 1964.

\bibitem{CDG}
C. \ Cohen-Tannoudji, J. \ Dupont-Roc and G. \ Grinberg,\
 {\it Atom-Photon Interactions, Basic Processes and Applications,}
John Wiley \& Sons, Inc., 1992.

\bibitem{Fei}
E.~L.\ Feinberg,\
 {\it A particle with non-equilibrium proper
field}, in: Problems of theoretical physics, Memorial volume to
Igor E. Tamm, Nauka, Moscow, 1972, pp. \ 134--149.


\bibitem{RS}
V.~A. \ Rubakov and M.~E.\ Shaposhnikov,
{\it  Elektroweak
baryon number non-conservation in the early Universe and in
high-enegry collisions}, Uspehi Fisicheskih nauk, \ 166
 (1996), 493--538.

\bibitem{WM}
D.~F.\ Walls and G.~J. \ Milburn, {\it Quantum
Optics}, Springer-Verlag,  New York, 1994.

\bibitem{Vol1}
I.~V. \ Volovich, {\it Models of Quantum Computers and
Decoherence Problem}, $~~~$\\
e-Print archive http://xxx.lanl.gov/
quant-ph /9902055;  Proc. of the International Conference
on Quantum Information, Meijo Univ. Nagoya, 4-8 Nov. 1997.

\bibitem{AAV}
L.\ Accardi, I.~Ya.\ Aref'eva and I.~V. \ Volovich,
{\it Non-Equilibrium Quantum Field Theory
and Entangled Commutation Relations},
e-Print archive http://xxx.lanl.gov/hep-th/9905035.


\bibitem{Bog1}
N.N. Bogoliubov,  {\it Problems of dynamical theory in statistical
physics}, Gostehizdat, Moscow, 1946.

\bibitem{vHo1}
L. \ van \ Hove, {\it Quantum mechanical perturbations giving
rise to a transport equation}, Physica, \ 21 (1955), 517--540.


\bibitem{PH}
I. Prigogine,  {\it Non-equilibrium statistical mechanics},
Pergamon, 1963.


\bibitem{AcKoVo}
L. \ Accardi, S.~V.\
Kozyrev, I.~V. \ Volovich,
{\it Dynamics of dissipative two--state systems in the
stochastic approximation},  Phys. Rev. \ A 57  (1997), 1111-1127;
e-Print archive http://xxx.lanl.gov/quant-ph/9706021.


\bibitem{AcLuVo}
L. \ Accardi, Y.~G. \ Lu and I.~V.\ Volovich,
{\it Quantum theory and its stochastic limit}, Springer, (in press)


\bibitem{AV5} I.~Ya.\ Aref'eva and I.~V. Volovich, 
{\it Quantum decoherence 
and higher order corrections to the large time 
exponential behaviour}, quant-ph/9906022.

\bibitem{Are}
I.~Ya.\  Aref'eva, \ Teor.\  Mat.\  Fis. {\it Renormalized
waves operators, II},
 \  15 (1973), 207-215.

\bibitem{BV}
N.~L.\  Balazs and and A. \ Voros,
{ \it The quantum Baker`s Transformation},
Ann. Phys., New York,  \ 190 (1989), 1--31.

\bibitem{TVS}
F.\ Toscano, R.~O. \ Vallejos and M. \ Saraceno,
 {\it Boundary
contributions to the semiclassical
traces of the baker`s map}, e-Print archive
http://xxx.lanl.gov/chao-dyn/9703005.

\bibitem{KH}
L.\ Kaplan and E.~J. \ Heller,
{\it Overcoming the Wall in the
Semiclassical Baker`s Map},
e-Print archive http://xxx.lanl.gov/chao-dyn/9809008.

\bibitem{SS}
A.~N. \ Soklakov and R.\ Schack,
{ \it Classical limit in terms of symbolic
dymanics for the quantum baker`s map} ,
e-Print archive http://xxx.lanl.gov/quant-ph/9908040.

\bibitem{IOV}
K.\ Inoue, M.\ Ohya and I.~V. \
Volovich, {\it Semiclassical Properties
and Chaos
Degree for the Quantum Baker`s Map} (in preparation).

\bibitem{AlbSeb}
S.\ Albeverio  and  P.\ Seba, {\it Wave chaos in
quantum systems with point interaction}, J. Stat. Phys.,
\  64 (1991), 369--383.

\bibitem{AMRV}
I.~Ya.\ Aref`eva, P.~B.\ Medvedev, O.~A. \ Rytchkov and
I.~V.\ Volovich, {\it Chaos in M(atrix) Theory} , Chaos,
Solitons and Fractals, \  10 (1999), 213--223.

\end{thebibliography}
\end{document}